\documentclass[runningheads]{llncs}
\usepackage[utf8]{inputenc}
\usepackage[T1]{fontenc}
\usepackage[usenames,dvipsnames,svgnames,x11names,table]{xcolor}
\usepackage{cite}
\usepackage{amsmath,amssymb,amsfonts}
\usepackage{algorithmic}
\usepackage{graphicx}
\usepackage{textcomp}
\usepackage{listings}
\usepackage{url}
\usepackage{pifont}
\usepackage[colorlinks=true,linkcolor=blue,urlcolor=blue,citecolor=blue,bookmarks=false]{hyperref}
\usepackage[abbreviations]{foreign}
\usepackage{enumitem}
\usepackage{booktabs}
\def\BibTeX{{\rm B\kern-.05em{\sc i\kern-.025em b}\kern-.08em
    T\kern-.1667em\lower.7ex\hbox{E}\kern-.125emX}}

\newboolean{showcomments}
\setboolean{showcomments}{true}
\ifthenelse{\boolean{showcomments}}
{ \newcommand{\mynote}[3]{
   \fbox{\bfseries\sffamily\scriptsize#1}
   {\small$\blacktriangleright$\textsf{\emph{\color{#3}{#2}}}$\blacktriangleleft$}}}
{ \newcommand{\mynote}[3]{}}

\newcommand{\xcite}[1]{}

\usepackage[font={small},skip=1pt,belowskip=3pt,labelfont=bf]{caption}
\usepackage{listings}
\definecolor{backcolour}{rgb}{0.95,0.95,0.92}
\lstdefinestyle{yaml} {
    columns=fixed,
    numbers=left,                   
    stepnumber=1,                   
    numbersep=6pt,                  
    numberstyle=\footnotesize\ttfamily,      
    xleftmargin=11pt,
	framextopmargin=1pt,
	aboveskip=0pt,
    identifierstyle=\color{black}\ttfamily,
    stringstyle=\color{purple}\ttfamily,
    keywords=[1]{platform, architecture, cpu, os,"passthru-devices",volumes},
    keywordstyle=[1]\color{blue}\ttfamily,
    keywords=[2]{features, hardware},
    keywordstyle=[2]\color{NavyBlue}\ttfamily,
    keywords=[3]{SGX, sse4, GPU},
    keywordstyle=[3]\color{blue!50!purple}\ttfamily
}

\begin{document}

\title{Analysis and Improvement of Heterogeneous Hardware Support in Docker Images}
\titlerunning{Heterogeneous Hardware Support in Docker Images}

\author{Panagiotis Gkikopoulos\inst{1}\orcidID{0000-0001-6436-8929} \and Valerio Schiavoni\inst{2}\orcidID{0000-0003-1493-6603} \and Josef Spillner\inst{1}\orcidID{0000-0002-5312-5996}}
\institute{
Zurich University of Applied Sciences, Switzerland,
\email{\{pang,spio\}@zhaw.ch}
\and
University of Neuch\^atel, Switzerland,
\email{valerio.schiavoni@unine.ch}
}

\maketitle

\begin{abstract}
\vspace{-10pt}
Docker images are used to distribute and deploy cloud-native applications in containerised form. A container engine runs them with separated privileges according to namespaces. Recent studies have investigated security vulnerabilities and runtime characteristics of Docker images.
In contrast, little is known about the extent of hardware-dependent features in them such as processor-specific trusted execution environments, graphics acceleration or extension boards.
This problem can be generalised to missing knowledge about the extent of any hardware-bound instructions within the images that may require elevated privileges.
We first conduct a systematic one-year evolution analysis of a sample of Docker images concerning their use of hardware-specific features.
To improve the state of technology, we contribute novel tools to manage such images.
Our heuristic hardware dependency detector and a hardware-aware Docker executor \textit{hdocker} give early warnings upon missing dependencies instead of leading to silent or untimely failures.
Our dataset and tools are released to the research community.
\footnote{\color{SeaGreen} Copyright  notice:  The  final  publication  is  available  at  Springer  via \url{https://doi.org/10.1007/978-3-030-78198-9_9}.}
\vspace{-10pt}
\end{abstract}


\vspace{-4pt}
\section{Introduction}\label{sec:introduction}
\vspace{-4pt}
Portability is a desirable property in cloud computing~\cite{petcu2011portability,di2014applications}\xcite{bozman2010cloud}.
Virtual machine images, container images and other executable artefacts ought to be flexibly deployed across clouds~\cite{DBLP:books/sp/aws14/BinzBKL14}, migrate from the cloud to the edge~\cite{DBLP:journals/csi/CarrascoDP20}\xcite{DBLP:conf/ucc/MurphyE18}, or be confined within differentiated cloud security zones \cite{DBLP:conf/wcnc/SchinianakisTMC19}.
Nevertheless, limitations to portability arise from the differences in the underlying heterogeneous hardware.
Apart from targeting the basic processor architecture, new cloud service paradigms such as accelerated computing or secure computing demand support for further hardware extensions abstracted as shared resources~\cite{DBLP:conf/hpdc/YehCC20}.
Any hardware thus requires support in the boot process (BIOS/EFI), operating system, virtualisation layers and application execution stacks, as well as in the applications and their constituent artefacts (\eg images).
Among the hardware features of interest to cloud application engineers are FPGAs, GPUs and nested virtualisation~\cite{DBLP:journals/integration/ShepovalovA20,DBLP:journals/dt/TarafdarELC18,DBLP:journals/jss/RenQDYS17}, as well as security-increasing hardware.

Secure (confidential) cloud computing is one of the emerging concepts affected by reduced portability: tenants offload computation to third-party untrusted providers without having to worry about security issues typically associated with cloud offloading~\cite{DBLP:journals/ieeesp/Herardian19}.
Over the years, this concept has been mapped to concrete secure architectures involving monitoring and intrusion detection~\cite{modi2013survey}, algorithmic proofs on data, trusted boot~\cite{johnson2018titan}, physical security including hardware security tokens and modules~\cite{DBLP:journals/iot/CoppolinoDMR19} and policy languages\xcite{DBLP:conf/ucc/KebbediesSBS15}.
Recently, hardware-based trusted execution environments (TEEs) enabled outsourcing the processing of confidential data with high privacy requirements~\cite{Brito2019} towards untrusted clouds.

Mapping this concept to hardware requires support for special CPU instructions as well as appropriate configuration of hypervisors and other execution parts.
Several vendor-specific instruction sets exist nowadays in processors available in the consumer market, as well as those offered by cloud providers~\cite{google-confidential,aws-confidential}.
Specifically, we mention Intel Software Guard Extensions (SGX)~\cite{costan2016intel}, AMD Secure Enhanced Virtualisation (SEV)~\cite{kaplan2016amd}, or ARM TrustZone~\cite{pinto2019demystifying,amacher2019performance}.

In most cloud environments, Docker containers become the standard \emph{de facto}, as application developers migrate towards integrated platform services, \eg IBM Code Engine\xcite{bluemix}, Amazon ECS\xcite{ec2cs}, Azure Container Service\xcite{awscs} and Google Container Engine\xcite{googlece}.
Research on heterogeneous hardware support for container execution has been limited to few aspects, and in particular has not covered TEEs.
For highly security- and privacy-sensitive software applications, such as those in the domains of e-health, banks or smart metering, the use of TEEs offers many benefits.
To drive their adoption, it is necessary to encapsulate the use in containerised form.
Yet container runtime support for heterogeneous hardware in public clouds is still lacking, in particular for secure execution.
This is surprising given that certain features are today available to providers of cloud infrastructure-as-a-service.
For instance, all major cloud providers offer instances with SGX-/SEV-capable CPUs, nevertheless access to consumers is either limited or disabled.
More specifically, Azure offers SGX/SEV capabilities with their Confidential Compute instances, while IBM or Alibaba have them enabled only for their bare metal instances.
AWS and Google do not provide SGX-enabled instances, as the feature is disabled in their systems' BIOS.\footnote{Google just recently announced SEV-enabled instances~\cite{google-confidential}, while AWS is introducing Nitro Enclaves, heavily inspired by Intel SGX~\cite{awsnitro}.}
In general, no cloud provider natively supports a container runtime that makes use of SGX.
We believe this situation is going to fundamentally change over the next few years, and in that perspective, the research focus will shift to provide efficient support in the application stacks and artefact management tools, nowadays largely container-based.

To allow a Docker container to utilise a hardware-specific feature of the host system, the relevant device driver or kernel module needs to be explicitly mounted to the container at runtime.
Some features, such as SGX, can be configured to run in simulation mode in the absence of supporting hardware.
Base images that offer the relevant development tools for these features are sometimes provided by the hardware vendors themselves (\eg, Nvidia Docker~\cite{nvidiadocker}), other companies (\eg, SCONE~\cite{10.5555/3026877.3026930,scontain}) or by members of the developer or research community.
For the SGX use-case, the Graphene Secure Container Environment (GSCE)~\cite{gsce} aims to wrap the standard Docker engine to allow standard images to run their applications within SGX enclaves, although this is in preliminary stages of development.

While our analysis shows that only few Docker images require specialised hardware at the time of writing, this is expected to change in the future.
Already now, the discoverability of hardware in the form of CPU architectures per tag is limited, leading to issues in mixed environments.
Making conscious use of the hardware features requires knowledge about them, which is presently absent from container images and handling tools and motivates our research.
By consciously leveraging specific hardware types, augmented images can be invaluable to certain sectors, \eg, Highly Information Sensitive Computing (HISC), High Performance Computing (HPC) and FPGA-hosted blockchain-as-a-service~\cite{DBLP:conf/sped/FlorinI19,DBLP:conf/cse/ChoLBK19}, enabling the integration of containers to their niche workloads.

The contributions of this paper are fourfold.
We introduce a \textit{systematic analysis} of the content of Docker images with emphasis on heterogeneous hardware, especially on TEEs for security and GPUs for acceleration.
In contrast to existing empirical works, our analysis follows a three-staged process: automated monitoring for metadata inspection, static image inspection, and manual runtime inspection.
We also contribute \textit{augmented metadata} that reflects the analysis results, and an improved \textit{hardware-aware Docker executor} \texttt{hdocker} that parses the metadata to increase upfront assessment about the likelihood of hardware-dependent execution failure.
We follow an \textit{open science} approach. Our tools and datasets are released and available from \url{https://doi.org/10.5281/zenodo.4531794}.

\textbf{Roadmap.} This paper is organized as follows.
In \S\ref{sec:background}, we provide a more comprehensive background on processor architectures and Docker container images.
\S\ref{sec:method} describes our three-step methodology.
We present our evaluation strategy and experimental results in \S\ref{sec:eval}.
In \S\ref{sec:discussion}, we present our proposed extensions.
We cover related work in \S\ref{sec:related}.
Finally, we conclude and present future work in \S\ref{sec:conclusion}.

\vspace{-4pt}
\section{Background}\label{sec:background}
\vspace{-4pt}
\subsection{Processors and Hardware Heterogeneity}

At the most fundamental level, hardware support means the ability to execute on a given processor architecture.
Public clouds typically include \texttt{amd64} (\texttt{x86-64}) and \texttt{arm64}, with minority occurrences of \texttt{i386}, \texttt{ppc64el} (IBM) or \texttt{mipsel} (Loongson).
However, discussions of heterogeneous hardware refer to a broader set of features, such as specific instruction sets.
When supported by BIOS/EFI, hypervisor and operating system, applications can typically rely on more than 100 unique CPU flags, \eg, \texttt{sse3} for streaming operations or \texttt{avx} for vectoring.
This also includes hardware-assisted TEEs, \ie Intel SGX and AMD SEV or hardware-assisted virtualisation technology (\ie, Intel VT-x, AMD-V, HVM, \etc).
Similarly, another group captures support for expansion devices or cards.
This became increasingly relevant in recent years, by the usage of GPUs for high-performance computing, machine learning and simulations.
We include in this group support for fast/low-latency disks (\emph{e.g.}, NVMe non-volatile memories) and network links (\emph{e.g.}, InfiniBand).

\vspace{-6pt}
\subsection{Docker Images}
\vspace{-4pt}
Since its introduction in 2013, Docker emerged as one the most popular cloud-native software artefacts and as such an integral part of modern DevOps operations
Several frameworks and development environments are now available as drop-in solutions using Docker, significantly reducing the overhead of both setting them up and orchestrating them.

Images can be distributed in two ways.
The first largely adopted method is using a container registry, the largest one being DockerHub.\footnote{Docker Hub: \url{https://hub.docker.com/}}
There are private registries, curated by specific vendors\footnote{Red Hat Registry: \url{http://quay.io}, Tenable: \url{http://tenable.io}},  which incorporate stricter standards, \ie built-in security scanning of all images motivated by several security issues found on public registries~\cite{10.1145/3029806.3029832}.
The second distribution method is to include the Dockerfile in a publicly accessible code repository (\eg, GitHub), for users to build the image themselves~\cite{7962382}.

\begin{figure}[!t]
    \centering
    \includegraphics[scale=0.7]{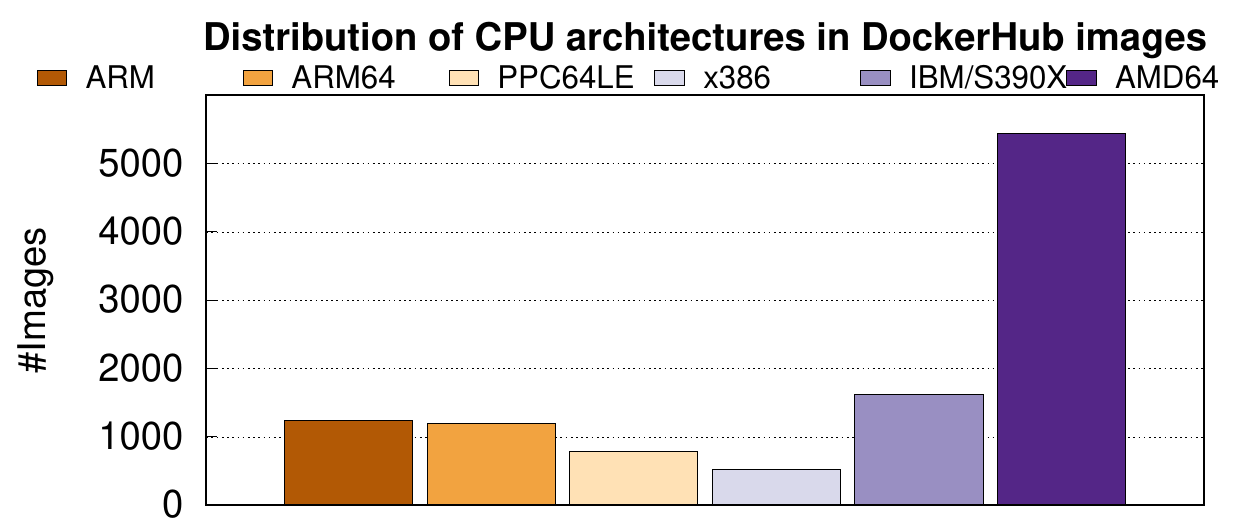}
    \caption{Distribution of images types in an $\sim$11'000 images non-random sample collected as of February 4$^{th}$, 2021. X-axis: processor architecture.}
    \label{fig:docker:archs}
	\vspace{-12pt}
\end{figure}

\begin{figure}[!t]
    \centering
    \includegraphics[scale=0.7, trim={10 0 0 150}]{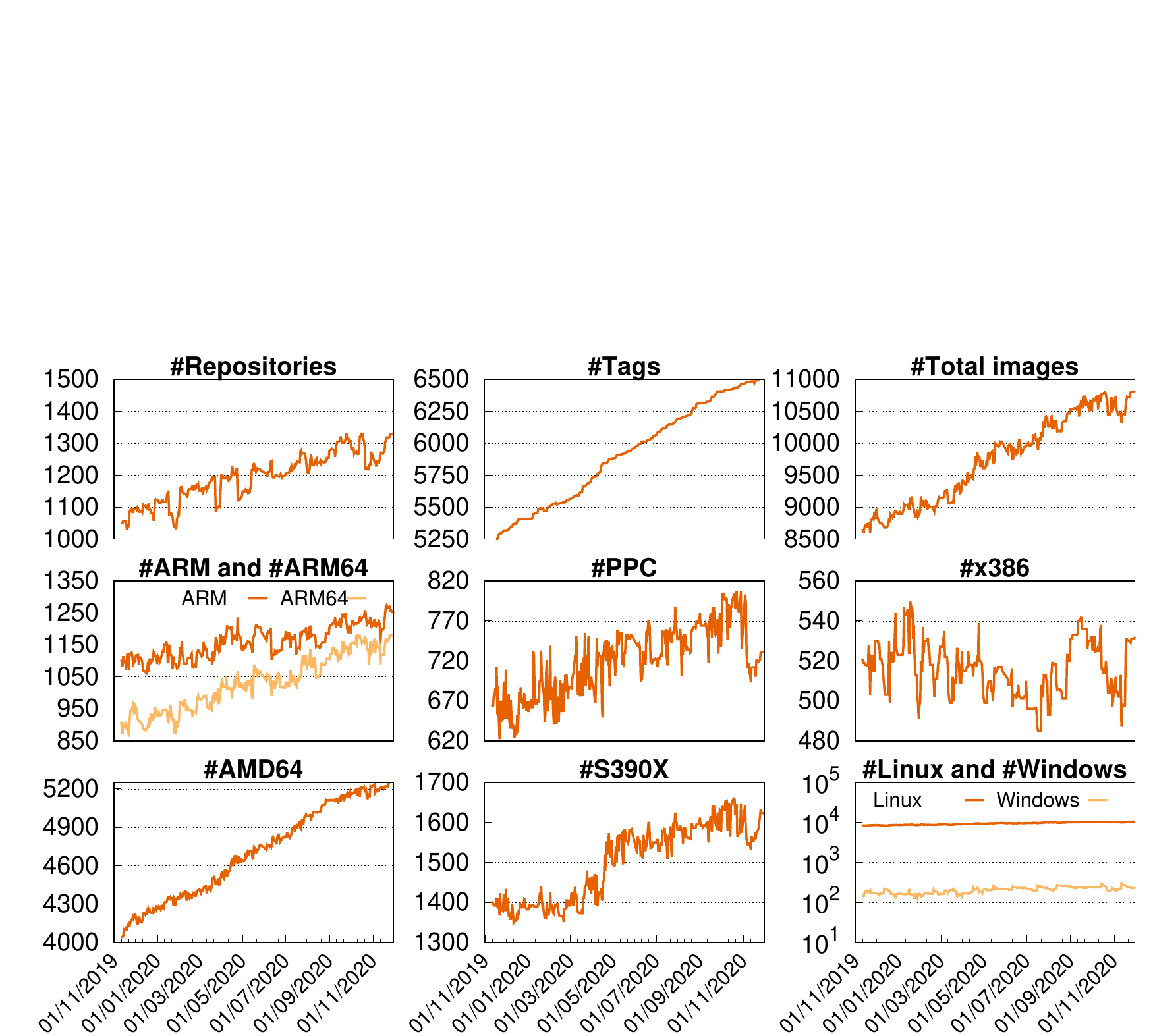}
    \caption{Monitoring metrics for Nov 11, 2019 to Feb 4, 2021}
    \label{fig:docker:trends}
\end{figure}

Docker has support for multiple CPU architectures, namely \texttt{x86}, \texttt{x86-64}, \texttt{ARM}, \texttt{ARM64}, \texttt{PowerPC 64 LE} and IBM's POWER and Z architectures.
We fetched and analyze a sample of the publicly accessible Docker images on DockerHub, constituted by approximately 11'000 images (as we detail later in \S\ref{sec:method}).
We report these results in Figure~\ref{fig:docker:archs} (also released to the research community, see Appendix~\ref{sec:data}).
Unsurprisingly, \texttt{x86-64} leads by a large margin with $\sim$48\% of the images.
Interestingly, IBM Z (\texttt{S390X} in Figure~\ref{fig:docker:archs}) is second most popular (with $\sim$15\%).
We observe that \texttt{ARM64} is the fastest trending image type, after \texttt{x86-64}, showing a 1.7\% monthly increase (Figure~\ref{fig:docker:trends}).
Docker has recently introduced an experimental multi-architecture builder in which several versions of an image, compatible with multiple processor architectures are built simultaneously, thus speeding up the process of introducing compatibility for different architectures.

Metadata available from the official registry provides basic overview metrics, such as supported architectures and OS versions, as well as the number of images and tags within a repository.
However, global statistics cannot currently be obtained from DockerHub, as the API endpoint for the global catalog metadata is disabled.
Utilization of Docker images in contexts that require specialized hardware is now an emergent use-case, varying from Nvidia GPUs to Intel SGX-compatible CPUs, though adoption appers to be at an early stage, judging from the number of repositories and their omission from the list of official images.
They require the host to have the correct hardware and driver support installed, and are thus limited in their reach to end users, but provide the ability to leverage Docker's advantages for applications that require specialized hardware features, such as secure execution and GPU-accelerated HPC.
Still, metadata does not reflect hardware dependencies, and that makes such repositories difficult to detect with automated means, which is why we searched by specific vendors (namely Nvidia and Intel) and hardware features (CUDA and SGX) heuristically to create the sample for the evaluation of hardware support.

\vspace{-6pt}
\section{Methodology}\label{sec:method}
\begin{figure}[!t]
    \centering
    \includegraphics[scale=0.6]{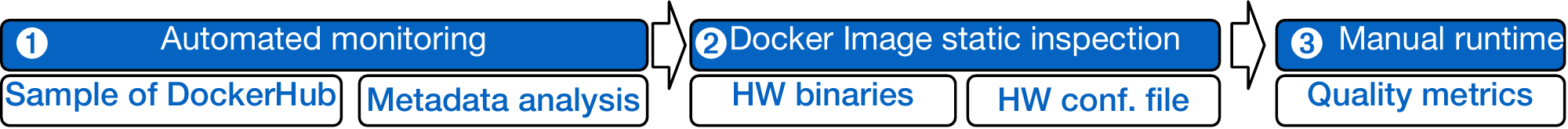}
    \caption{Our methodology to gather and validate DockerHub's image data/metadata.}
    \label{fig:method}
	\vspace{-16pt}
\end{figure}

\vspace{-4pt}
The methodology followed to assess the hardware support in Docker images is divided into three steps.
The first one (Figure~\ref{fig:method}-\ding{202}) consists of automatic monitoring of the main DockerHub repository to retrieve a representative sample of the most used images.
The metadata from the sample images is then stored and analysed for information on any specific hardware they support or require.
The second step (Figure~\ref{fig:method}-\ding{203}) dives deeper into the images and attempts to find any traces of hardware-specific binaries or configuration files.
Finally, the third one (Figure~\ref{fig:method}-\ding{204}) performs basic manual runtime testing of a targeted selection of Docker images from DockerHub that use specific hardware features, to assess their quality and maturity.

The automated sampling of DockerHub metadata uses a two-tier approach.
The search starts with the official Docker images, collected in a private \emph{library} repository.
The metadata from this repository is collected and added to the dataset.
For each entry in the resulting data, we assume each image to correspond to an organisation, which maintains its own repository.
For example, the image \texttt{nginx} is controlled and maintained by a corresponding \texttt{nginx} organization.
We thus repeat the search for each of these possible organisations, adding the data returned (if any) to the dataset.
This process essentially creates a shallow family tree of two levels, stemming from the official images. This serves to effectively increase the sample size by an order of magnitude,
granting us a higher number of images to investigate. This selection process is scripted and both it and the metadata analysis runs nightly on a scheduled monitoring setup.

Compared to typical random sampling, this method has the advantage of specifically targeting repositories within DockerHub that are considered of high-quality/high-profile, as their maintainers are either verified or certified publishers.
It should be noted that the goal of this sampling is to feed the next steps with potential targets, not to obtain global metadata statistics for Docker Hub as a whole. As such, the statistics are meant as an indication of hardware support in the official and related repositories and do not represent a study of Docker Hub as an ecosystem.

Currently, the metadata provided by DockerHub's API does not provide any insight on whether or not an image requires specific hardware to be present on the host in order to function.
A possible \emph{hardware features} field could rectify this.
In the present state of the metadata, the only way to select images with specific hardware requirements beyond CPU architecture is to rely on the web front-end's text search function, a rather cumbersome and unreliable approach.
For example, a search of the keyword 'SGX' with the intention of discovering images that utilise Intel SGX, will also yield images submitted by users with the characters 'sgx' in their username.
Naturally, such a search will exclude any images that use the hardware but don't include it in their name or description.

However, due to the relatively small number of images using targeted hardware technology, it is still possible to hand-pick a sample and assess them heuristically for typical quality metrics, such as compatibility, documentation and best practices.
For the sake of illustration, we restrict the choice of the target technologies to: \emph{(1)} Intel SGX, as a secondary processor architecture feature with high significance for security-conscious use cases, and \emph{(2)} support for Nvidia GPUs, arguably the most common example of heterogeneous computing power.
For the former, adoption is still rather limited.
We believe this to change as SGX drivers have just been up-streamed into the mainline Linux kernel\footnote{42nd rev \url{https://lore.kernel.org/lkml/20201214114200.GD26358@zn.tnic/}}.
For the latter, adoption is more wide-spread, with many individual user-contributed repositories, so the search is narrowed down to the images contributed by the hardware vendor themselves, and restricted to static analysis.

\vspace{-6pt}

\section{Evaluation}\label{sec:eval}
\begin{figure}[!t]
    \centering
    \includegraphics[scale=0.6]{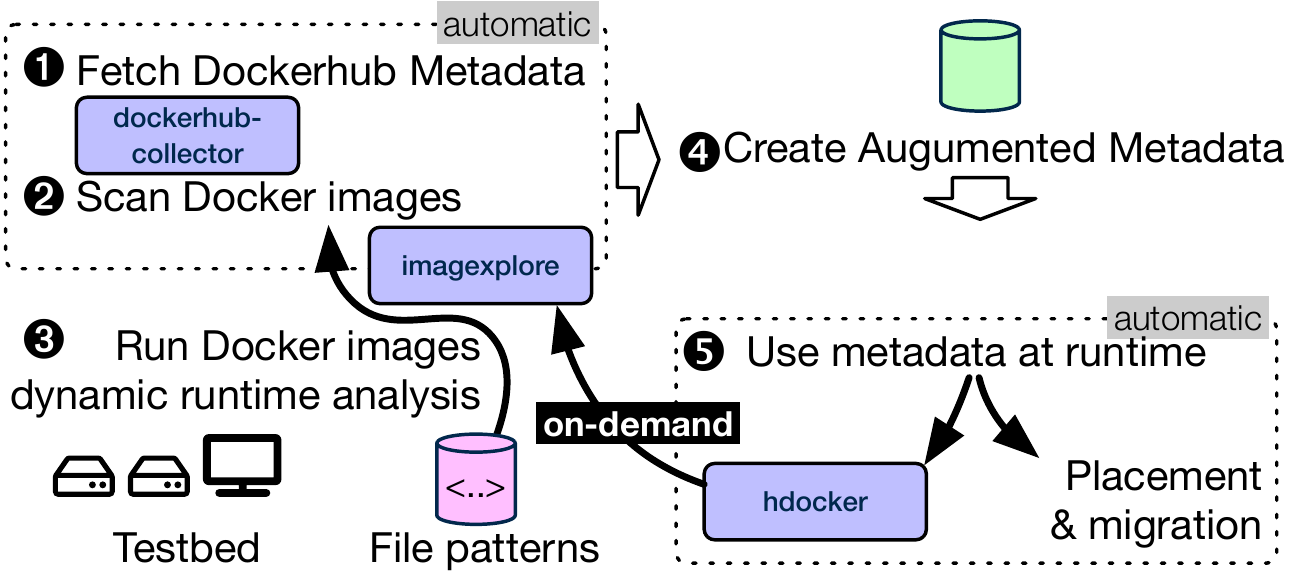}
    \caption{Data-driven process: from experiment (steps \ding{202},\ding{203},\ding{204}) to augmented metadata usage (\ding{205},\ding{206}); \texttt{dockerhub-collector}, \texttt{imagexplore}, and \texttt{hdocker} are open-source.}
    \label{fig:process}
\end{figure}

\vspace{-6pt}
We ran three separate experiments, one for each of the aforementioned methodic steps.
Figure~\ref{fig:process} gives an overview about the experiments, the resulting augmented metadata and its use in specific
container and cloud management tools.

\textbf{Automated monitoring.}
We setup automated long-term monitoring and tracking of Docker container images via the global Microservices Artefact Observatory.\footnote{MAO: \url{https://mao-mao-research.github.io/}}
The observatory creates execution schedules for each registered monitoring tool, and shares these tools with other nodes within a federated cluster architecture for highly reliable observation and metadata retrieval.
The \sloppy{\texttt{dockerhub-collector}} tool periodically requests DockerHub's public API to retrieve metadata on the 'library' repository.
This repository is unique: it is curated by DockerHub itself and contains 'official images', \ie the most useful, popular and high-quality images available.
Using this data as a starting point, we identify 'sister repositories', \ie attempt to see if the developer of each image maintains their own repository, for which we request metadata. This selection process is integrated into the tool and automated.
It increases the sample from the 164 repositories of the official set to $\sim$1200 repositories.

The continuous metadata collection has been running on each machine in the cluster daily, starting on November 11, 2019.
The data snapshots obtained by each machine are compared for inconsistencies
and then aggregated into a single dataset to analyse the hardware trends.
The images with SGX and Nvidia support are selected manually in irregular intervals, based on a search of the repository for the specific hardware feature as explained before. The final lists describing the samples,
after filtering out duplicates, images without any version and irrelevant repositories, are included in the dataset to ease reproducibility.
Thus, we curated multiple lists of images as February 2021 snapshot for the evaluation: full library and associated images ($\sim$11'000), strict library subset (164), SGX (67) and Nvidia (36).
We applied static scanning to the last two, and dynamic assessment to the last.

\textbf{Static image inspection.} To gain further insights into the hardware-related aspects of container images, we developed \texttt{imagexplore}, a tool to uncompresses all layers of an image and recursively search for the occurrence of certain files, directory and content patterns.
These come from a knowledge base created with file glob patterns as well as text and binary symbols hinting at the presence of hardware-specific features.
Matches result from the installation of the SGX toolkits, Nvidia device drivers, USB devices or other hardware-related software leaving traces in the system.
For example, an SGX base image needs to contain the SGX SDK.
Thus, toolkit files can be detected in a predictable manner to confirm the presence of the SDK.
Extending the knowledge base with additional patterns for different hardware devices enables the detection of images that utilize them, giving information on potential deployment issues due to hardware incompatibility.
Evidently, this experiment leads to suspicions but not always to verified cases due to false positives.
Hence, our method encompasses the manual verification of any edge cases. Furthermore, the scanning encompasses all files in the \texttt{/dev} file system, as those are not supposed to be part of any container image but their accidental addition may reveal further insights.
This scanning phase is performed for all the automatically retrieved library and SGX images.

\textbf{Manual runtime inspection.} Beyond the static testing, we conduct dynamic runtime invocation on the SGX images to find out the actual use of any TEE-related operation.
This test is run on an SGX machine using an Intel Core i7-8650U.
We categorised the SGX images into \emph{(1)} base images and \emph{(2)} those for applications or tools.
We assessed them on the presence of documentation, instructions and the ability to initialise on an SGX-compatible machine.
It is important at this stage to define what 'initialise' means in this context.
For an application or tool, this is trivial: we expect some program output or the successful initialisation of a service.
For a base image, we expect to setup a functional development environment when we run the image in interactive mode.
This was achieved by inspecting the image to ensure the SDK and relevant dependencies are present.
In the case of curated images, code samples are provided in the documentation that can be used to test the image.
Due to the small number of SGX images, we allowed a minimal troubleshooting, which we define being able to run the images by just following the instructions (if any).
In case of errors, we see if it is straightforward to resolve, \eg mounting a file in a volume or specifying a command-line argument.
If not, the image is deemed as not working, although its usability mainly depends on incomplete (or lack of) documentation.
The minimal configuration possible for an image without instructions or other forms of documentation is to run it with no options besides mounting the SGX-related virtual devices from the host (\ie, \texttt{/dev/isgx} and \texttt{/dev/mei0}).
For base images, this proved sufficient on the majority of occasions, as the images contained all the development tools they needed except for the hardware driver component.

\vspace{-8pt}
\subsection{Results}
\vspace{-2pt}

\begin{figure}[!t]
    \centering
    \includegraphics[scale=0.7]{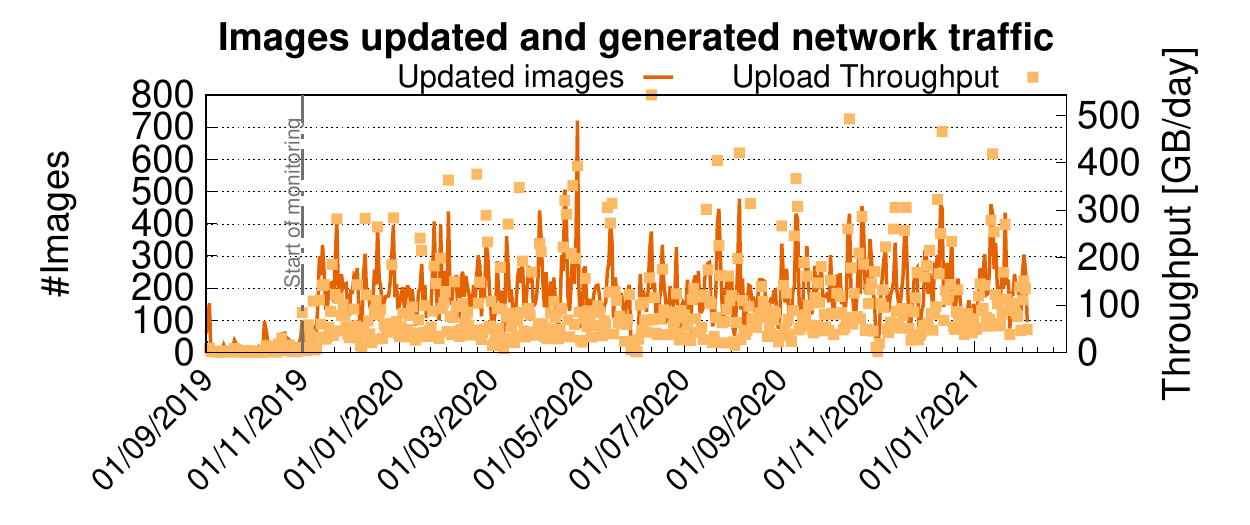}
    \caption{Images updated and data uploaded from Sept'19 until Feb'21; \texttt{y2axis} (right side): inbound network traffic originated by updates of the images on the \texttt{y1axis}.}
    \label{fig:my_label3}
\end{figure}

\begin{figure}[!t]
    \centering
    \includegraphics[scale=0.7, trim={0 0 0 20pt}]{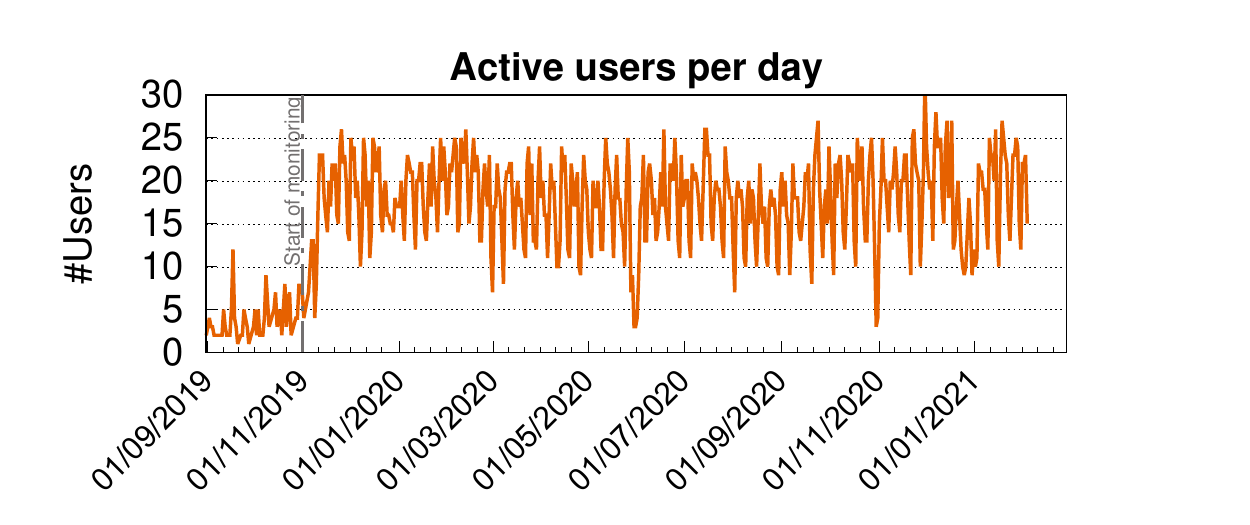}
    \caption{Active users by day from Sept'19 until Feb'21.}
    \label{fig:my_label4}
\end{figure}

\textbf{Automated monitoring results.}
The automated monitoring of Docker Hub's public metadata gives the basic overview of processor architecture and operating system support discussed above.
While not novel in itself, we further consider the amount of activity on the chosen sample during the same time period.
Figure~\ref{fig:my_label3} and Figure~\ref{fig:my_label4} present respectively the amount of image updates and total data volume per day, as well as the number of unique usernames that appeared in each day's data, extracted from the same dataset. The goal of this observation is to ensure that our sample does not come from an inactive subset of Docker Hub, and thus reflects recent trends.
In the observed period, the total amount of data pushed to these repositories was $\sim$23TB.

\textbf{Static image inspection results.} We run the \texttt{imagexplore} tool on three sets of container images: \emph{(1)} 67 SGX candidates, \emph{(2)} 36 Nvidia candidates, and \emph{(3)} 164 official library images as subset of the sample whose metadata we continuously retrieved over multiple months.
Due to the absence of a \texttt{:latest} tag or non-public images, not all pulls succeed, thus these numbers are slightly reduced especially for less maintained non-official images for which non-default tags have therefore been recorded.
All layers of all images are unpacked on disk for the analysis.
We expect the tool to indicate 100\% SGX and Nvidia dependencies in the first two sets (minus false positives), and 0\% any hardware dependency in the third set (minus false negatives).
Table~\ref{tab:imagexplore} compares the results.
Among the SGX candidates, four use the name coincidentally but do not contain any SGX-related code.
Similarly, among the suspected Nvidia images, four contain code to run the Kubernetes-based EGX AI platform that is offered by Nvidia but do not contain any hardware-related code themselves.
Among the library images, three are operating system distributions with support for USB devices that are successfully detected, whereas the vast remainder are pure application containers.
The average scanning time per image already present locally is 7.0~s for library images, 15.1~s for Nvidia images and 30.3~s for SGX images on a server with Xeon E3-1270 CPU @ 3.80GHz.
The main contributor to the significant difference is that most library images are rather tidy base images whereas especially the SGX images contain many layers with thousands of added files.

\begin{table}[!t]
\centering
\caption{Automated static analysis of the SGX, Nvidia and library container samples.\label{tab:imagexplore}}
\setlength{\aboverulesep}{0pt}
\setlength{\belowrulesep}{0pt}
\rowcolors{1}{gray!10}{gray!0}
\begin{tabular}{llll}
\rowcolor{gray!25}
\toprule
							& \textbf{SGX}	& \textbf{Nvidia} & \textbf{Library}	\\
\rowcolor{gray!25}			& \textbf{images}& \textbf{images} & \textbf{images}	\\
			\hline
Count					& 67		& 36		& 164			\\
Successfully retrieved			& 62 (93\%)	& 14 (39\%) 	& 156 (95\%)		\\
Cumulative disk size			& 61 GB		& 15 GB		& 75 GB			\\
Hardware dep detection correctness	& 58 (94\%)	& 10 (71\%)	& 152 (98\%)		\\
Detection time				& 1876.7 s	& 211.0 s	& 1080.4 s		\\
Exceptions				& False pos.	& False pos.	& False neg.		\\
Excepted images				& sgx-django,	& egx-*,	& odoo, crux,		\\
					& sgx-bootstrap,& eac		& sourcemage		\\
					& sgx-cosmos,	&		&			\\
					& payload-ethash&		&			\\ \hline
\end{tabular}
\vspace{-10pt}
\end{table}

Concerning the \texttt{/dev} file system, the official images show better quality compared to the SGX images.
However, in both sets, a lot of superfluous files were found.
The Nvidia images are quality-wise in between with still a few of such files.
Most of them mirror the standard device files added automatically by Docker, whereas almost as many consist of various other files to access audio devices, RAM
disks, shared memory and other hardware.
We suspect careless creation of images by snapshotting running container instances without proper post-execution cleanup of device files is the main reason.
This hypothesis is supported by the occasional leftover of temporary files (\texttt{/tmp}, \texttt{/var/tmp}) in both sets of images.
Table~\ref{tab:devstats} compares the device file findings, with the optimal metric being 0 in all rows.

\begin{table}[!t]
\centering
\caption{Device files in the SGX and library container samples.\label{tab:devstats}}
\setlength{\aboverulesep}{0pt}
\setlength{\belowrulesep}{0pt}
\rowcolors{1}{gray!10}{gray!0}
\begin{tabular}{llll}
	\rowcolor{gray!25}
	\toprule
							& \textbf{SGX}	& \textbf{Nvidia} & \textbf{Library}	\\
	\rowcolor{gray!25}		& \textbf{images}& \textbf{images} & \textbf{images}	\\
			\hline
Average /dev layers	& 2.1		& 1.0		& 1.0		\\
Multiple /dev layers	& 10 (15\%)	& 0 (0\%)	& 8 (5\%)	\\
Shadow /dev files	& 27..36 (53\%)	& 1..2 (14\%)	& 6..11 (7\%)	\\
Further /dev files	& 27 (40\%)	& 1 (7\%)	& 3..8 (5\%)	\\
\end{tabular}
\vspace{-8pt}
\end{table}

\textbf{Manual runtime inspection results.}
The mentioned categories yield 33 base images (49.25\% of total), 29 apps/tools (43.28\%) and 5 images (7.46\%) whose functionality could not be determined due to lack of documentation.
The latter ones are excluded from the percentage calculations that follow.
As for quality metrics, 47 of the images (75.81\%) initialised successfully when tested and only 14 of them (22.58\%) had any documentation or instructions.

Considering Docker Hub's limitations on making images private, the bulk of the undocumented images are likely experiments that were not meant to be shared.
Even within these 67 images, a certain degree of deduplication was possible.
Upon manual inspection 28 of the images were derivatives of others in the same set, of which 18, based on their functionality, were duplicated of earlier versions of other images.
Examples include several variations of the SGX base image, which provides the SGX SDK and Platform Software (PSW) and an example 'hello world' application, cementing the observation that many images available were simply the results of users experimenting with the technology, or following tutorials.
The lack of documentation and the amount of 'hello world' duplicates in the sample is a clear indication that the majority of images that use the SGX technology are experimental, and thus adoption of the technology is in its infancy on the Docker ecosystem.

In terms of observed errors among the 15 failures, there were image initialisation errors for 4 images and an enclave initialisation for 1.
Others were application-specific errors.
It is important to mention that as with any heuristic evaluation, the testing methodology can affect the final result and as such, the 15 failures have to be treated as an upper bound.
As a secondary experiment, we attempted to run the tools that ran successfully, but without mounting the SGX devices.
For them, 8 out of 15 ran successfully in simulation mode, with only two 'hello world' applications and an SGX-specific stress testing tool failing due to enclave initialisation errors.
The 4 remaining failures were instances where the output of the program made it difficult to understand if initialisation was successful, \ie possibly false-negatives.

The high success rate of the base images, even those with no documentation, also has a simple explanation. These images usually only include the SGX dependencies (the SDK and PSW) and typically some language specific dependencies for development. Thus, generically running them interactively by simply mounting the SGX device drivers is expected to yield a high degree of success. The spreadsheet used to track this heuristic evaluation is available in the publicly released dataset.

Images with support for Nvidia GPUs are far more numerous, potentially up to 1411 image repositories in our latest snapshot.
We note that the company maintains 2 distinct organisation accounts (\texttt{nvidia} and \texttt{nvidiak8s}, respectively with 22 and 12 repositories), the latter focused on Kubernetes-related images.
A common characteristic of both is the almost complete absence of documentation within Docker Hub itself.
Instead, all relevant documentation is available in Nvidia's Github repositories, where the source code for these images can also be found.
The absence of links from the Docker Hub repositories to the ones in Github however, means that the only reliable source of information on these images and their usage is Nvidia's Github. This is by no means an oddity, as typically Docker Hub is used to store images for public distribution and serves as a medium for users that wish to use a specific tool without building from source code manually, one of the main advantages of Docker. In that context, only having minimal documentation on Docker Hub (usually related to differences in using the image compared to running a tool natively) makes sense, as the main source code repo is responsible for housing the full documentation.

\begin{figure}[t!]
\noindent\begin{minipage}[t]{.4\columnwidth}
\begin{lstlisting}[style=yaml, caption=Hardware features extensions.,  captionpos=t,label=yaml:features,basicstyle=\small]
{
  "platform": {
    "architecture": "amd64",
    "os": "linux",
    "features": [
      "sse4"
    ]}}
\end{lstlisting}
\end{minipage}\hfill
\hspace{-15pt}
\begin{minipage}[t]{.55\columnwidth}
\begin{lstlisting}[style=yaml, caption=Support for heterogeneous hardware sources.,  captionpos=t,label=yaml:features:hw,basicstyle=\small]
{
  "platform": {
    "cpu": {
      "architecture": "amd64",
      "features": {
        "SGX": "supported",
        "sse4": "required"
      }
    },
    "hardware": {
      "GPU": "nVidia"
    },
    "os": "linux"
  }}
\end{lstlisting}
\end{minipage}
\vspace{-23pt}
\end{figure}

\vspace{-8pt}
\subsection{Augmented Metadata}
\vspace{-8pt}
As mentioned earlier, Docker images built to utilize specialized hardware can benefit from augmented metadata to include hardware requirements.
Such images are still a relatively small niche within the ecosystem.
However, additional metadata can benefit users attempting to deploy them, as a clearer picture can be obtained by the Docker client on whether the image can be expected to function in the target system.
To this end, we propose hardware-specific fields be added to the metadata of Docker registry entries.
Furthermore, with Docker Hub's new rate-limiting policy (\ie, 100 image pulls every 6 hours), pulling an image simply for the sake of inspecting is problematic for free-tier users: with better metadata, operators can make informed decisions about image pulls ahead of time.
Note that part of the additional information we propose to include is contained in the experimental image manifest v2, schema 2 of the Docker registry API specification\cite{dockermanifest} (see Listing~\ref{yaml:features}, \texttt{features} element).
Evidently, this implementation has some limitations.
First and foremost, the image manifests are not available via the public API.
As a consequence, one must pull the image and inspect it manually, a process that is hard to automate and requires storage space and network bandwidth for then potentially non-executable images. For instance, \texttt{docker run} will currently pull an AMD64 image on ARM before complaining about incompatible binary formats.
Secondly, in terms of hardware, only CPU architectures and instruction sets are considered.
Consequently, one cannot identify if an image uses GPUs or FPGAs from its manifest alone.
Finally, CPU architecture variants and features are written to the manifest manually but cannot be verified during validation.
This leaves the inclusion of this information at the discretion of the developer and the validation up to the end user.

Thus, we propose an extension of the platform information that distinguishes CPU architecture and features, as well as other hardware devices supported.
Listing~\ref{yaml:features:hw} shows an example in augmented JSON format that serves as downward-compatible replacement for the current Docker registry metadata.
We argue that it is important to distinguish a hardware feature as \texttt{supported}, \eg, its functionality can be emulated/simulated if the hardware component is missing), or \texttt{required}, \eg, the container will fail without it.

Moreover, specific hardware support from the host needs to be passed through to the running container.
Hence, rather than replicating pass-through parameters in all scripts and composition documents (Listing~\ref{yaml:features:pass}), we argue for tightly binding them with the container metadata.
Similar to how permissions are confirmed by users of mobile phone applications, administrators can interactively approve applications based on displayed host device access permissions.
In addition, some containers require volumes to be mounted, or access to the Docker daemon socket, to be indicated in the extended metadata structure.

\begin{figure}[t]
\begin{lstlisting}[style=yaml, caption=Pass-through support.,  captionpos=t,label=yaml:features:pass,basicstyle=\small]
{
  "platform" as defined above
  ...
  "passthru-devices": [
    "/dev/kvm", "/dev/net/tun", "/dev/bus"
  ],
  "volumes": [
    "/run/secrets", "/var/run/docker.sock"
  ]}
\end{lstlisting}
\vspace{-10pt}
\end{figure}

\vspace{-2pt}
\subsection{Improved Container Image Tooling}
\vspace{-2pt}

Finally, we introduce \texttt{hdocker}, a hardware-aware wrapper around the \texttt{docker run} command, to perform several tasks that streamline the use of Docker containers across heterogeneous hardware architectures.
Specifically, \texttt{hdocker} can:

\begin{itemize}[noitemsep,topsep=0pt,parsep=0pt,partopsep=0pt]
\item Report on the availability of tags per architecture and architectures per tag.
\item Check whether CPU flags and hardware devices are present and activated, beyond the standard CPU architecture and operating system.
\item Check whether device files and volumes are appropriately declared to be mounted in the invocation parameters.
\item Inform the user with clear actionable advice in case of unsupported features.
\item Automate the analysis (using imagexplore) and decisions if asked to do so and if possible.
\end{itemize}

The \texttt{hdocker} tool is metadata-driven, as it parses the augmented metadata curated from our experiments.
Unless cached locally, it downloads all container image layers, runs \texttt{imagexplore} over all images, and store the resulting JSON metadata files. If the registry uses our augmented schema to retain this information, end users would not have to perform the image pull and analytics to obtain the hardware requirements of the image and \texttt{hdocker} could configure the hardware pass-through automatically.

Beyond the designed and implemented \texttt{hdocker} CLI, we also envision further tools to assess the metadata in a similar way. Among them are autonomous container migration tools that work as target node filters in heterogenous computing continuums (\ie cloud/edge/fog/IoT or osmotic settings \cite{DBLP:journals/computer/VillariFDRJR19}).

\vspace{-10pt}
\section{Discussion}\label{sec:discussion}
\vspace{-6pt}
While Docker images that support or require specialised hardware still comprise a very small share of the total available images, we argue that their significance to industries that make use of heterogeneous or specialised hardware justifies communicating support for such features in an image's public metadata.

Our proposed support is nevertheless not a simple addition to the metadata schema.
There are certain requirements for making its adoption meaningful.

\emph{Docker CLI support.} The hardware features would have to be specified by the user through the Docker CLI during a build, or detected by the Docker builder. 
The latter is harder to implement, but would enforce the hardware specification, whereas developers would neglect specifying the hardware requirements.

\emph{Dockerd support.} When attempting to run an image with specific hardware requirements that are not met, the Docker daemon would need to alert the user of the issue, rather than \texttt{hdocker}.

\emph{Docker Registry support.} The hardware-related metadata would have to be part of the public Docker registry API, so that end-users can make informed decisions on the images they deploy ahead of time. The need for this is accentuated by the rate limit introduced by Docker Hub recently, forcing operators to be mindful of image pulls they make.

\vspace{-10pt}
\section{Related Work}\label{sec:related}
\vspace{-6pt}
Analyzing and understanding large container datasets offers substantial insights on current trends among practitioners and DevOps engineers' best practices.
In~\cite{DBLP:conf/cluster/ZhaoTAARSW0B19}, authors produce a large-scale dataset of more than 450'000 Docker images by crawling Docker Hub, downloading and decompressing images and analysing their contents. 
The dataset is populated with file-level metrics such as compression ratios, file types and sizes, and content-level metrics, \emph{i.e.},  programming languages, compressed files and database systems. 
Despite rich metadata including deduplication metrics, the underlying OS or architecture features are ignored. 
Moreover, the dataset is not made available, forcing the research community and ourselves to implement, deploy and run our own crawler to get complementary insights.
While our automated monitoring is on a smaller scale, we produce and release a curated dataset that provides historical metrics by monitoring at regular intervals over a longer period of time. 
Our dataset also serves as an example of the type of insights that could be obtained by applying our methods to other repositories, such as those maintained by specific organisations.

Approximate Concrete Execution~\cite{ace} applies binary analysis to container images used in serverless platforms such as Google Cloud Run and AWS Fargate.
It uses function-level fingerprinting to identify security problems in an early stage, based on a database associating fingerprints to such issues. 
The same technique could be used to fingerprint hardware dependencies. 
For simplicity, we look at container images at a coarse-grained level -- if any script or application therein depends on a specific hardware feature, no matter if this is run on startup or just placed accidentally in the image, we consider the dependency to exist.

Several container orchestration tools have been recently proposed specifically for heterogeneous hardware clusters.
However, they lack the necessary detection and expression of hardware dependencies within the container images or composed applications.
DRAPS~\cite{DBLP:journals/corr/abs-1805-08598} focuses on resource allocation, and thus in quantitative differences in CPU performance, amount of RAM and network/block device I/O latency, whereas our work investigates hardware differences at a deeper and qualitative level. 
Our work is complementary and necessary to further improve heterogeneous cloud application orchestration with upfront predictability about the likelihood of failure.

Dockemu extends the network simulator NS-3 with portable containers to cover IoT scenarios~\cite{DBLP:conf/simultech/PortabalesN18}. 
It supports heterogeneity in nodes and links between them, but does not cover hardware differences in detail. 
We anticipate the use of our results for building better heterogeneous cloud simulators and emulators.

The metadata expressivity limitation of Docker Hub and privately deployed container image registries have been widely covered by researchers. 
ConHub~\cite{DBLP:conf/cikm/TianPT16} is an extended metadata management system for Docker images on top of a relational database. 
It supports CQL queries and user-defined metadata, letting users manage hardware-dependent images when importing augmented metadata, similar to the ones we propose.
Similarly, Docker2RDF~\cite{ayed2017docker2rdf} and DockerFinder~\cite{brogi2017dockerfinder} offer queryable endpoints that will benefit from richer descriptions including on hardware restrictions at pull time. 
While we focus on hardware features, studies exist~\cite{7962382} to analyse image content in terms of programming languages, runtimes, security vulnerabilities and reproducible builds, leading to further metadata.

\vspace{-12pt}
\section{Conclusion and Future Work}\label{sec:conclusion}
\vspace{-8pt}
Our multi-method analysis on Docker container images identified limitations on the handling of heterogeneous hardware, an increasingly relevant matter in typical container environments such as differentiated cloud services.
We released a new DockerHub dataset on the metadata of $\sim$11'000 Docker images over the period of more than one year.
We offer insights on available and used hardware features, \eg related to security (SGX) and machine learning (Nvidia).
We validated initial search findings with static analysis, and cross-validated those with runtime testing.
Further, we contributed new proposals to augment the container image metadata with hardware, device and volume information as well as tools to exploit the knowledge at runtime for more user-friendly application handling.
Overall, we observed a solid growth of images and tags across architectures, with the exception of
\texttt{x386}, whilst a low number of non-toy, non-base images exploiting specific hardware beyond the basic CPU architecture.
As future work, we aim at extending container image quality benchmarks with hardware details, and look further into enclaved and isolated application execution in the cloud.

\section*{Data and Code}\label{sec:data}

Metadata, analysis data, tools and code for reproducibility are made publicly available at \url{https://doi.org/10.5281/zenodo.4531794}.

\bibliographystyle{splncs04}
\bibliography{paper}

\end{document}